\documentclass{article}

\usepackage{arxiv}

\usepackage[utf8]{inputenc} %
\usepackage[T1]{fontenc}    %
\usepackage{hyperref}       %
\usepackage{url}            %
\usepackage{booktabs}       %
\usepackage{amsfonts}       %
\usepackage{nicefrac}       %
\usepackage{microtype}      %
\usepackage{lipsum}		%
\usepackage{graphicx}
\usepackage[authoryear]{natbib}
\usepackage{amsmath}
\usepackage{mathrsfs}
\usepackage{multirow}
\usepackage{bbm}

\DeclareMathOperator{\sign}{sign}

\title{A probabilistic view on rupture predictability:\\all earthquakes evolve similarly}

\author{Jannes Münchmeyer$^{1, 2, *}$, Ulf Leser$^{2}$, Frederik Tilmann$^{1, 3}$\\
$^1$ Deutsches GeoForschungsZentrum GFZ, Potsdam, Germany\\
$^2$ Institut für Informatik, Humboldt-Universität zu Berlin, Berlin, Germany\\
$^3$ Institut für geologische Wissenschaften, Freie Universität Berlin, Berlin, Germany\\
$^*$ To whom correspondence should be addressed: \url{munchmej@gfz-potsdam.de}
}

\renewcommand{\headeright}{}
\renewcommand{\undertitle}{}
\renewcommand{\shorttitle}{A probabilistic view on rupture predictability}

\begin{document}
\maketitle

\begin{abstract}
 Ruptures of the largest earthquakes can last between a few seconds and several minutes.
 An early assessment of the final earthquake size is essential for early warning systems.
 However, it is still unclear when in the rupture history this final size can be predicted.
 Here we introduce a probabilistic view of rupture evolution - how likely is the event to become large - allowing for a clear and well-founded answer with implications for earthquake physics and early warning.
 We apply our approach to real time magnitude estimation based on either moment rate functions or broadband teleseismic P arrivals.
 In both cases, we find strong and principled evidence against early rupture predictability because differentiation between differently sized ruptures only occurs once half of the rupture has been observed.
 Even then, it is impossible to foresee future asperities.
 Our results hint towards a universal initiation behavior for small and large ruptures.
\end{abstract}

\section{Introduction}

It is a longstanding question at which time during an earthquake rupture its final size can be constrained.
Answering this question would have direct implications for early warning systems \citep{allenEarthquakeEarlyWarning2019} and would provide insights into the underlying physical processes.
Accordingly, its formulation spanned a series of studies over the last decades.
However, so far results have been contradictory - some argue for early predictability, others against. 

A common theory implying predictability is the preslip model \citep{ellsworthSeismicEvidenceEarthquake1995}, in which failure starts aseismically until the process reaches a critical size and becomes unstable.
Here, the final moment of the earthquake might be derivable at the event onset time from properties of the nucleation zone, i.e., its size or its magnitude of slip.
Other models also suggest early predictability, but only after several seconds.
For example, \citet{melgarSystematicObservationsSlip2017} argue that ruptures of large events propagate as self-healing pulses, and that pulse properties allow identification of very large events after $\sim$15~s.
Support for such theories has been provided by the analysis of, e.g., waveform onsets \citep{ellsworthSeismicEvidenceEarthquake1995}, moment rate functions \citep{danreEarthquakesEarthquakesPatterns2019a}, and early ground motion parameters \citep{colombelliEarlyRuptureSignals2020}.

The opposing hypothesis, often termed cascade model \citep{ellsworthSeismicEvidenceEarthquake1995}, suggests a universal initiation behavior: small and large earthquakes start identically and are differentiated only after the peak moment release, which occurs approximately at half of the rupture duration.
Rupture evolution is controlled by heterogeneous local conditions, such as the pre-event stress distribution or the presence of mechanical barriers.
Studies supporting this theory also analyzed properties like moment rate functions \citep{meierHiddenSimplicitySubduction2017}, waveform onsets \citep{ideFrequentObservationsIdentical2019}, or peak displacement \citep{trugmanPeakGroundDisplacement2019}.

While reaching contradicting conclusions, predictability studies often follow the same principle: analyzing correspondences between earthquake size and real time observables \citep{ellsworthSeismicEvidenceEarthquake1995, danreEarthquakesEarthquakesPatterns2019a, colombelliEarlyRuptureSignals2020, meierHiddenSimplicitySubduction2017, ideFrequentObservationsIdentical2019, trugmanPeakGroundDisplacement2019}.
Earthquake size is commonly quantified by seismic moment/moment magnitude, as large, high quality catalogs thereof are openly available \citep{ekstromGlobalCMTProject2012}.
A common practice is calculating parametric fits between magnitude and observables, and assessing at which time they become significant using standard deviations \citep{melgarSystematicObservationsSlip2017, danreEarthquakesEarthquakesPatterns2019a, colombelliEarlyRuptureSignals2020, meierHiddenSimplicitySubduction2017, olsonDeterministicNatureEarthquake2005, nodaScalingRelationEarthquake2016, zolloEarthquakeMagnitudeEstimation2006}.
However, this point-estimator approach hides the residual distribution and thereby potentially obscures distinct modes of rupture predictability, especially when distributions are non-Gaussian (see Figure \ref{fig:probabilistic_errors} for a toy example illustrating the importance of this restriction).
Notably, exactly such distributions occur for real observables; examples can be found in \citet[their Fig. 3, the observable is the dominant period of the initial 4 s of the P wave]{olsonDeterministicNatureEarthquake2005} or \citet[their Fig. 5, the observable is based on early P displacement waveform]{nodaScalingRelationEarthquake2016}.

\section{A probabilistic framework for rupture predictability}

\begin{figure}
 \includegraphics[width=\textwidth]{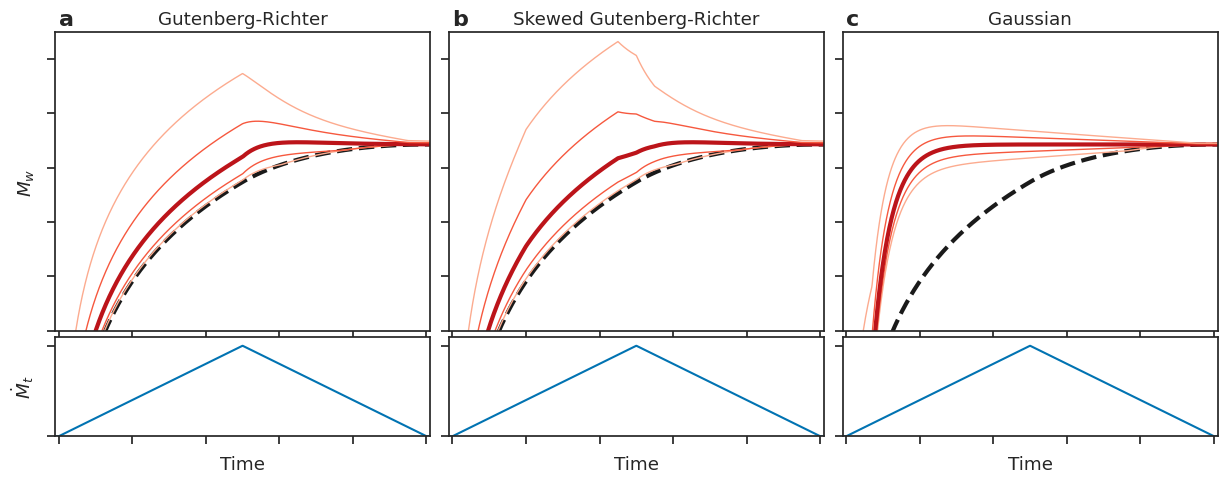}
 \caption{Magnitude estimate development for three different predictability models: Gutenberg-Richter (GR) (not predictable during growth phase), skewed GR (total magnitude not point-predictable, but information gain with respect to prior already during growth phase of rupture) and Gaussian (predictable). For each model, we use the same hypothetical event with prototypical triangular moment rate functions. The prototypical source time function is meant to represent the first order moment release history; for predictable models to be viable, second order features of the moment rate function would differ between smaller and larger events. Predictive distributions are visualized by their 0.05, 0.2, 0.5, 0.8, 0.95 quantiles over time. The cumulative moment release is indicated by the dashed black line. \textbf{a} In the GR case, the prediction follows a GR distribution above the moment released so far. Only after the peak moment release, the prediction quickly transforms into a Gaussian, although with a decreasing GR portion that relates to the possibility of future asperities. \textbf{b} The skewed GR case behaves similarly to the GR case, but the distribution is skewed towards higher magnitudes, i.e., from early on, it is more likely for the event to become large. \textbf{c} For the Gaussian case, the magnitude can be determined early on with small error that decreases further over time. No quantitative x and y labels are provided to highlight the prototypical character of the figure. A cross section view of the three different options at fixed time is shown in Figure \ref{fig:pmo_shapes_mixture}a.}
 \label{fig:models_over_time}
\end{figure}

We argue that a rigorous, probabilistic approach can overcome this issue.
To this end, we interpret the magnitude $M$ of an event as a random variable and introduce a stochastic process $(O_t)_{t \in \mathbb{R}}$, the observables at time $t$.
$t=0$ identifies the event onset.
The observables $(O_t)_{t \in \mathbb{R}}$ can be any information, as long as $O_t$ only describes the event until $t$, e.g., waveforms up to $(\text{P travel time} + t)$.

Events with magnitude $M_1 \neq M_2$ differ at time $t$ if the conditional distributions $\mathbb{P}(O_t|M_1)$ and $\mathbb{P}(O_t|M_2)$ differ.
However, while describing $\mathbb{P}(O_t|M)$ for scalar $O_t$ is feasible, it becomes intractable for higher dimensional $O_t$.
Furthermore, for early warning the objective is estimating $M$ from $O_t$ and not vice versa.
Therefore, we analyze $\mathbb{P}(M|O_t)$, directly investigating to what degree the observables constrain the magnitude.
While this type of analysis has been conducted for peak ground displacement, where \citet{meierHiddenSimplicitySubduction2017} considered $\mathbb{P}(O_t|M)$ and \citet{trugmanPeakGroundDisplacement2019} analyzed both $\mathbb{P}(O_t|M)$ and $\mathbb{P}(M|O_t)$, an analysis for higher dimensional observables is still missing.
This leaves many promising observables unexplored, e.g., seismic waveforms.

There are two distinct aspects of rupture predictability: (i) the future development of the current asperity and (ii) the probability of further asperities to rupture.
Figure \ref{fig:models_over_time}a shows an example of $\mathbb{P}(M|O_t)$ with no predictability in the growing rupture, as suggested, e.g., by \citet{meierHiddenSimplicitySubduction2017}.
Before the peak moment release, the distribution equals a Gutenberg-Richter (GR) distribution with lower bound at the currently released moment, accounting for both aspects.
After the peak, the distribution becomes Gaussian (i), with a decreasing GR component accounting for potential future asperities (ii).
Figure \ref{fig:models_over_time}b shows a skewed GR case: the magnitude cannot be pinpointed, but from early on the event is more likely to become large than the marginal GR distribution.
Skewed GR distributions might occur, e.g., in slip pulse models \citep{melgarSystematicObservationsSlip2017}, where pulse properties define the likelihood of the rupture to arrest soon.
Figure \ref{fig:models_over_time}c shows the predictable case: the magnitude can be pinpointed early and uncertainties decrease steadily, implying correct assessment of both aspects.

The different evolutions of $\mathbb{P}(M|O_t)$ have consequences for early warning: a shifted tail for $\mathbb{P}(M|O_t)$ shifts the estimated distribution of ground shaking and possibly the warning decision.
However, several previous results do not allow a clear distinction of the presented cases.
For example, multiple studies \citep{ideFrequentObservationsIdentical2019, abercrombieLocalObservationsOnset1994, kilbInitialSubevent19941999, moriInitialRuptureEarthquakes1996} reported that for most large events, small events with similar onsets exist.
While this rules out the predictable case, events might still differ strongly in their likelihood of becoming large.

For practical analysis, $\mathbb{P}(M|O_t)$ needs to be derived from observed samples $\{(M^i, O^i_t)\}_{i = 1, \dots, n} \sim_{iid} \mathbb{P}(M, O_t)$.
As direct description is infeasible for high dimensional $O_t$, we propose to instead use a variational approximation $\mathbb{P}_\theta(M|O_t) \approx \mathbb{P}(M|O_t)$ where parameters $\theta$ are learned to fit $\mathbb{P}(M|O_t)$ using samples $\{(M^i, O^i_t)\}_{i = 1, \dots, n}$ \citep{gneitingStrictlyProperScoring2007}.
Specifically, we suggest to parameterize $\mathbb{P}_\theta(M|O_t)$ using neural networks with Gaussian mixture outputs \citep{bishop1994mixture}, because both neural networks and Gaussian mixtures (Figure \ref{fig:pmo_shapes_mixture}b) have universal approximator properties, making them particularly well suited for our case \citep{cybenko1989approximation, bengio2017deep}.
This enables us to obtain probabilistic magnitude estimates, while not being restricted to single dimensional observables.
Notably, this approach can be applied directly to any type of observables, simply by designing an appropriate neural network.

\section{Predictions from moment rate functions}

\begin{figure}
 \includegraphics[width=\textwidth]{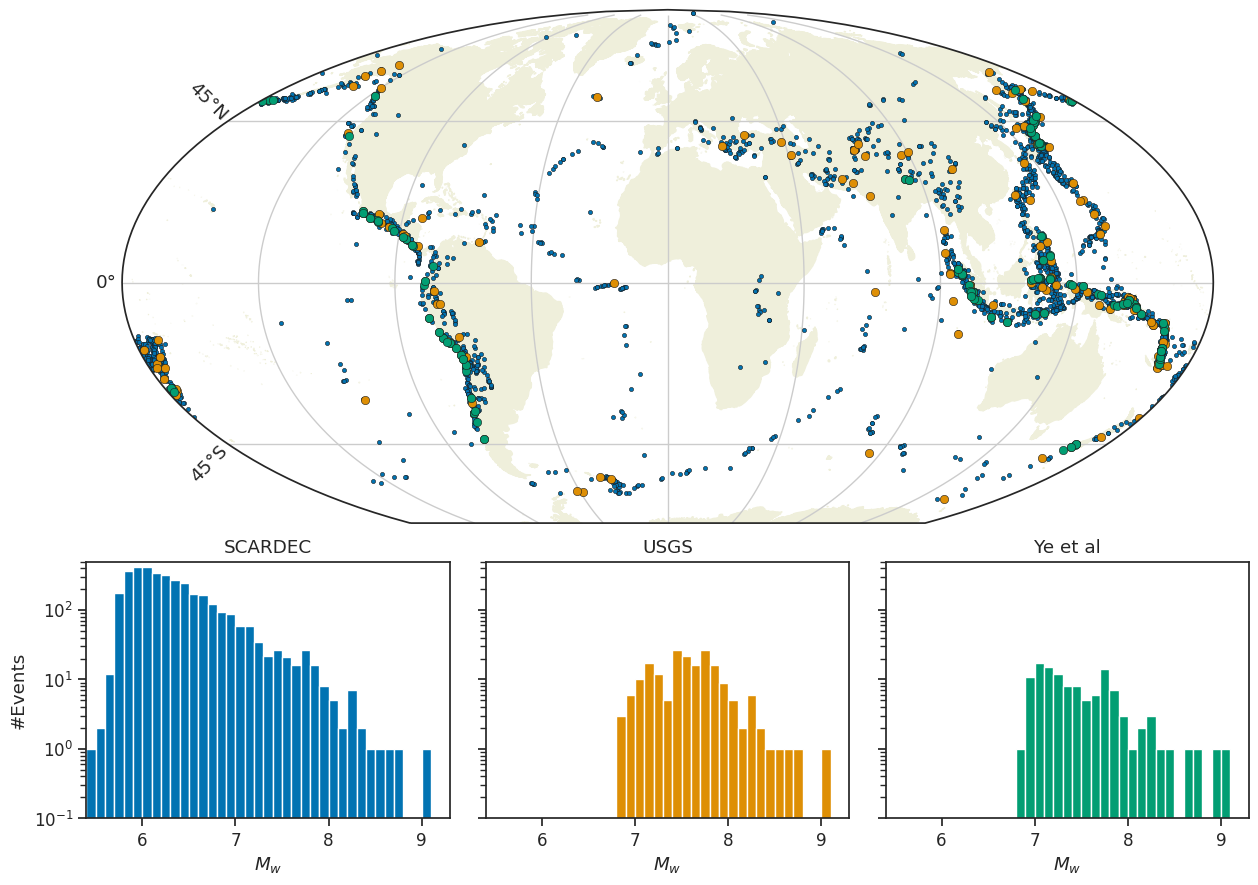}
 \caption{Distribution of events and histograms for magnitude distribution for the three STF datasets. The events are color coded by their dataset. Ye et al is plotted on top of USGS, on top of SCARDEC. This might lead to few events not being visible due to overlaps.}
 \label{fig:stf_map}
\end{figure}

\begin{figure}
 \includegraphics[width=\textwidth]{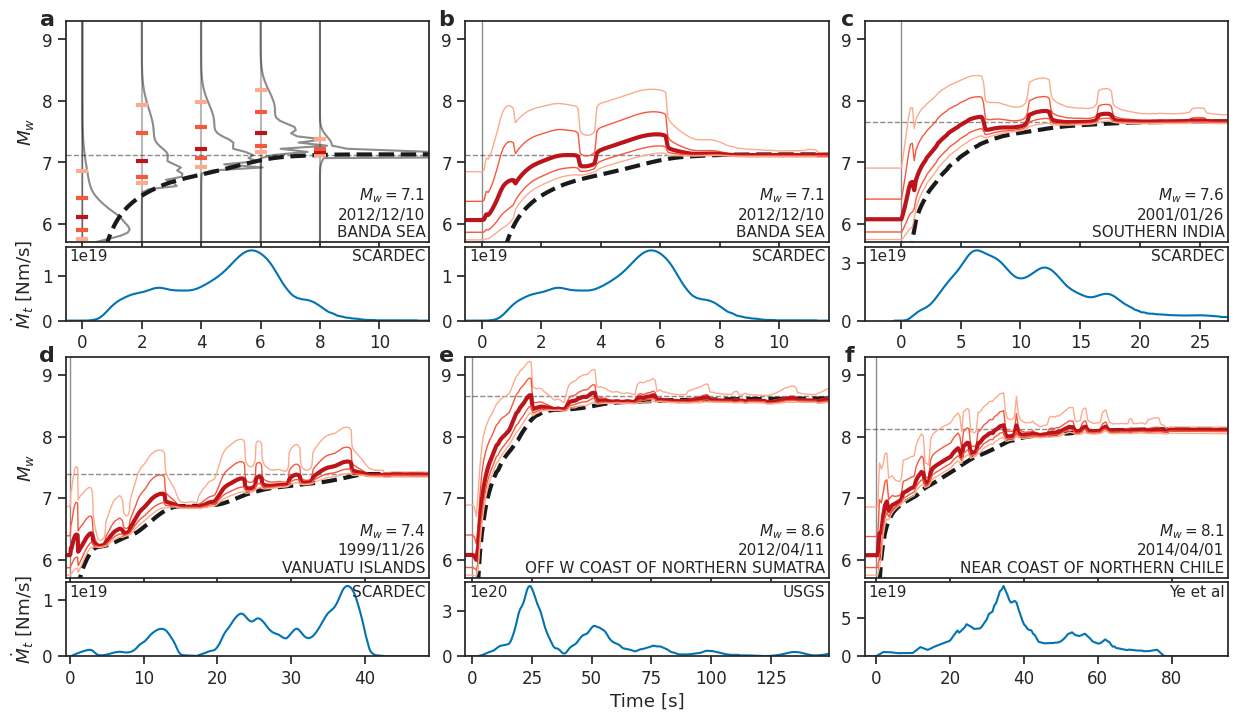}
 \caption{\textbf{a} Probability density functions (PDFs) calculated from the STF model just before onset, and at 2, 4, 6 and 8 s after onset. Colored ticks on the PDFs indicate 0.05, 0.2, 0.5, 0.8, 0.95 quantiles.
 \textbf{b-f} Example predictions from the STF model visualized by the 0.05, 0.2, 0.5, 0.8, 0.95 quantiles over time. \textbf{b} shows the same event as \textbf{a}. The lower right gives information on the event. The black dashed line shows the magnitude equivalent to the moment released so far, i.e., the trivial lower bound.
 The bottom plots show the STFs used for prediction. The annotation in the upper right indicates the STF database used.}
 \label{fig:results_stf_anecdotal}
\end{figure}

We first apply this framework to source time functions (STFs), also known as moment rate functions, a commonly used observable in predictability studies \citep{danreEarthquakesEarthquakesPatterns2019a, meierHiddenSimplicitySubduction2017}.
For our analysis we use three STF databases: SCARDEC (3514 events, $5.4 \leq M_w \leq 9.1$) \citep{valleeNewDatabaseSource2016} and the ones from USGS \citep{hayesFiniteKinematicRupture2017} (190 events, $6.8 \leq M_w \leq 9.1$) and \citet{yeRuptureCharacteristicsMajor2016} (119 events, $6.8 \leq M_w \leq 9.1$).
The geographic and magnitude distributions for all three datasets are shown in Figure \ref{fig:stf_map}.
The three STF databases were generated using two different methodologies.
SCARDEC uses a point source approximation and conducts a constrained deconvolution of body waves.
In contrast to SCARDEC, USGS and \citet{yeRuptureCharacteristicsMajor2016} calculate finite fault solutions from both body and surface waves assuming constant rupture velocity within each event.
As the spatial extent of the source is modelled, the STFs generally represent more high frequency details than the SCARDEC ones.
On the other hand, the SCARDEC method is applicable to smaller events that can not be processed with the finite-fault inversion schemes.
Further details on the methodologies of the STF datasets are provided in Text S\ref{sec:artifacts}.

As neural network model for the prediction of total moment magnitude based on (partial) source time functions, we use a simple multi-layer perceptron.
As input we use five observables derived from the source time function at time $t$: (1) cumulative moment $M_t$; (2) current moment rate $\dot{M}_t$; (3) average moment rate $\frac{1}{t} M_t$; (4) peak moment rate $\max_{\tau \leq t} \dot{M}_\tau$; (5) current moment acceleration $\ddot{M}_t$.
We use features instead of full STFs to avoid the danger of overfitting due to the high dimensionality of time series but low numbers of training examples.
Still, these features describe the STFs in sufficient detail to represent the observables considered in most previous STF based predictability studies \citep{meierHiddenSimplicitySubduction2017, melgarCharacterizingLargeEarthquakes2019}.
We train the model on SCARDEC, as it is the largest of the datasets, with further results from models trained on the USGS datasets available in Figure \ref{fig:results_stf_usgs}.
For training we use a ten fold cross validation scheme.
We use the continuous ranked probability score as loss, as its optimization behaviour is more favorable in face of highly skewed underlying distributions than for log-likelihood.
Further details on the model and training procedure are provided in Text S\ref{sec:stf_training}.

For qualitative insights into the predictions and as a basis for interpreting the average results, we visualize a few representative examples (Figure \ref{fig:results_stf_anecdotal}).
In all cases, the sign of the moment acceleration largely defines the anticipated potential for growth: positive acceleration, i.e., the growth phase, indicates high growth potential, negative acceleration low potential.
Furthermore, the higher the current moment release is, the higher the growth potential.
This results from the STF's smoothness: at high moment rates it will likely take longer to arrest than at low rates.
Notably, the model does not predict future asperities within a multiple asperities rupture (Figure \ref{fig:results_stf_anecdotal}d, e); for times after the peak of the moment rate function has been passed, the model expects a steady decay.
Once the moment rate approaches zero, the estimated further growth is low (e.g. Figure \ref{fig:results_stf_anecdotal}d at $15$~s, \ref{fig:results_stf_anecdotal}e at $40$~s).
If moment release accelerates again, the  model immediately expects another asperity to break and higher growth potential is inferred yet again. 
These effects lead to sudden changes of the PDF at local maxima and minima of the STF (e.g. Figure \ref{fig:results_stf_anecdotal}d at $20$~s, \ref{fig:results_stf_anecdotal}e at $25$~s).

\begin{figure}
 \includegraphics[width=\textwidth]{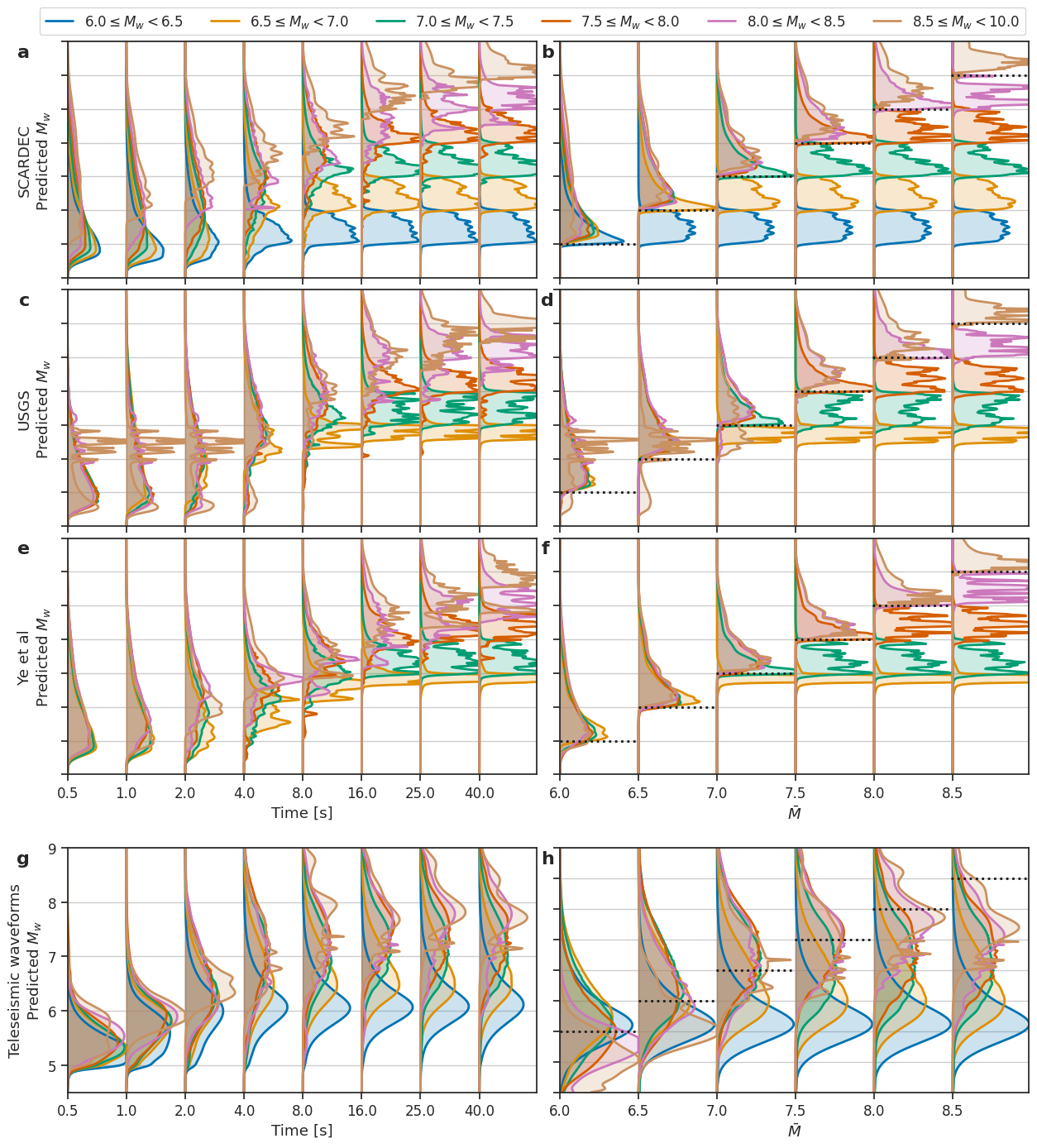}
 \caption{Average predicted PDFs based on STFs (\textbf{a-f}) and teleseismic waveforms (\textbf{g-h}) grouped by magnitude bin.
 Left column shows results at time $t$ after onset ($\mathbb{P}(M|O_t)$), right column after cumulative moment equals magnitude $\bar{M}$ ($\mathbb{P}(M|O_{\bar{M}})$).
 The STF model has been trained on the SCARDEC dataset and evaluated on each STF dataset.
 See Figure \ref{fig:results_stf_usgs} for STF results from a neural network trained with the USGS dataset.
 PDFs were truncated in visualization to avoid overlap between different times/base magnitudes.
 Black dotted lines in \textbf{b, d, f, g} indicate the current base magnitude.
 The apparent skew between buckets in panel \textbf{b} and prediction difference in \textbf{g} for $\bar{M} = 6.0$ likely results from SCARDEC processing artifacts (\ref{sec:artifacts}).
 For determining $t_{\bar{M}}$ in \textbf{h} we used the SCARDEC dataset. See Figure \ref{fig:results_team_datasets} for plots with the other STF datasets. Events differ between panels \textbf{g} and \textbf{h}: \textbf{h} only includes those events present in both the teleseismic dataset and SCARDEC ($\sim$3,500 events) and \textbf{g} all of the former ($\sim$38,000 events).
 }
 \label{fig:results_combined}
\end{figure}

For a systematic analysis, we average $\mathbb{P}(M|O_t)$ by magnitude buckets (Figure \ref{fig:results_combined}a, c, e).
For the datasets using finite fault solutions (Figure \ref{fig:results_combined}c, e), during the first 2~s of the STF, the predicted distributions are mostly identical across buckets.
Afterwards, the buckets split up over time: $M_w=6.5$ to $7.0$ at $\sim$2~s, $M_w=7.0$ to $7.5$ at $\sim$8~s, $M_w=7.5$ to $8.0$ at $\sim$16~s, $M_w=8.0$ to $8.5$ at 25--40~s.
These times match typical half-durations of events in these magnitude ranges \citep{gombergReconsideringEarthquakeScaling2016}.

SCARDEC (Figure \ref{fig:results_combined}a) shows similar splitting over time, but exhibits an apparent skew in the early predictions: higher magnitude buckets exhibit higher likelihood for becoming large.
Furthermore, lower bounds for the highest magnitude buckets (brown, purple), are higher than for the remaining buckets already after 1~s.
Similarly, the SCARDEC examples in Figure \ref{fig:results_stf_anecdotal}b-d show high predictions within the first 2~s and abruptly fall afterwards.
We attribute this apparent predictability to artifacts of the SCARDEC processing, in particular uncertainties in onset timing and the point source approximation (\ref{sec:artifacts}).
Additionally, we trained the model with the much smaller USGS dataset. The results lack the apparent early predictability, consistent with this explanation (Figure \ref{fig:results_stf_usgs}).

Predictions at a fixed time $t$ after onset describe both moment release until $t$ and future development, with only the later being relevant for predictability.
To isolate this aspect, we define $\mathbb{P}(M|O_{\bar{M}}) = \mathbb{P}(M|O_{t_{\bar{M}}})$, where $t_{\bar{M}} = \sup_t \{ M(t) \leq \bar{M} \}$ is the time when the cumulative moment release equals $\bar{M}$.
When analyzing $\mathbb{P}(M|O_{\bar{M}})$, all three datasets exhibit the same trends (Figure \ref{fig:results_combined}b, d, f).
All magnitude buckets with lower bounds at least $\bar{M} + 0.5$ show nearly identical predictions: a sharp increase in likelihood from $\bar{M}$ to $\sim \bar{M} + 0.2$ and an exponential tail.
$\bar{M} + 0.2$ represents roughly twice the seismic moment of $\bar{M}$ and, due to the symmetry of STFs \citep{meierHiddenSimplicitySubduction2017}, half the event duration.
For buckets with lower bound equal to $\bar{M}$, peak likelihood occurs around $\bar{M}$, again with exponential tails.
The decay is steeper for these buckets, as most events are already past the peak and substantial future growth can therefore only result from future asperities, but not from further growth of the current one.
The results are independent of the faulting mechanism (Figures \ref{fig:results_stf_eqtypes_scardec}, \ref{fig:results_stf_eqtypes_usgs}, \ref{fig:results_stf_eqtypes_yeetal}).
The systematic analysis therefore confirms the hypothesis that the final magnitude can only be assessed after the peak of the STF has been passed and that the rupture of further asperities cannot be anticipated.

\section{Predictions from teleseismic P arrivals}

STFs have limited temporal resolution %
such that high frequency details indicative of future rupture development might be hidden.
To resolve this issue, we apply our approach to teleseismic P arrival waveforms, which in contrast to STFs contain full spectral information up to $\sim 1$~Hz.
As neural network we adapted TEAM-LM \citep{munchmeyerEarthquakeMagnitudeLocation2021} and apply it to a catalog of $\sim$35,000 events with nearly 750,000 manually labeled first P arrivals (Figure \ref{fig:telemag_map}).

STFs have limited temporal resolution, giving only a low-pass filtered view of the source process.
Consequently, potential higher frequency details indicative of future rupture development might be hidden.
To resolve this issue, we apply our approach to teleseismic P arrival waveforms, which in contrast to STFs contain full spectral information up to $\sim 1$~Hz, above which they will be hidden by attenuation.
We collated a dataset of $\sim 35,000$ event with $\sim 750,000$ manually labelled first P arrivals.
As neural network we adapted TEAM-LM \citep{munchmeyerEarthquakeMagnitudeLocation2021}.
TEAM-LM consists of a combination of convolutional layers, a transformer network, and a mixture density output and predicts the event magnitude directly from the seismic waveforms at a flexible set of input stations.
Further details on the dataset and model are provided in appendix \ref{sec:predictability:teleseismic_data}.

Compared to the STF model, predictions $\mathbb{P}(M|O_t)$ and $\mathbb{P}(M|O_{\bar{M}})$ show higher uncertainties and systematic underestimation of the largest magnitudes at all times (Figure \ref{fig:results_combined}g).
Higher uncertainties result from the fact that assessing magnitude from waveforms is harder than from STFs, whereas underestimation can be attributed to data sparsity \citep{munchmeyerEarthquakeMagnitudeLocation2021}.
Due to the higher model uncertainties, all tails look rather like exponentially modified Gaussians than exponential distributions, as for the STF case.
We note that the apparent lower uncertainty for the highest magnitude buckets compared to the lower magnitude buckets results from the number of samples in each bucket: with fewer samples available, the result gets less smooth but also less wide.
Nonetheless, the general trends are highly similar to the results from the STF analysis before.
Early predictions ($t \leq 2$~s) are indistinguishable, except for the bin $M_w = 6.0$ to $6.5$, where event durations are often $<4$~s.
Bins split over time, similar to the STF model, although with higher overlap in predictions between bins.
Splits occur around 4~s for $M_w=6.5$ to $7.0$, 8~s $M_w=7.0$ to $7.5$, and 16 to 25~s for both $M_w=7.5$ to $8.0$ and $M_w=8.0$ to $8.5$, again representing typical event half-durations.

As for $\mathbb{P}(M|O_t)$, predictions for $\mathbb{P}(M|O_{\bar{M}})$ exhibit similar behaviour to the ones from the STF model (Figure \ref{fig:results_combined}h).
While $\bar{M}$ is considerably below the final magnitude, the predictions are indistinguishable between the buckets.
Splitting of buckets occurs slightly later than for the STF model, i.e., clear differences only become apparent once $\bar{M}$ exceeds the upper bound of the bucket.
This likely results from the higher uncertainties.
We therefore argue, that assessment is likely still possible from the moment peak onward, again up to potential further asperities.

\section{Comparison with previous results}
\label{sec:related}

Our results find no predictability from both STF and teleseismic waveforms.
These observations seem contradictory to several previous result.
We discuss potential reasons for some studies below.
\citet{melgarCharacterizingLargeEarthquakes2019} found differences in moment acceleration for earthquakes of different size.
As our STF model has access to the acceleration parameter investigated in this study, we would expect to be able to reproduce this effect.
However, later analysis demonstrated that these results were caused by a sampling bias, which our study confirms \citep{meierApparentEarthquakeRupture2020}.

\citet{danreEarthquakesEarthquakesPatterns2019a} analysed STFs as well, in this case by decomposing them into subevents, and also found predictability.
Large events exhibited higher moment in early subevents and in addition showed higher complexity, i.e., had more subevents.
We suspect that their different conclusion might also result from the SCARDEC processing, which hides small subevents within large earthquakes and thereby makes the first identifiable subevent within a large event comparably larger.
For a further description of artifacts in the SCARDEC dataset, we refer to Text S\ref{sec:artifacts}.

\citet{melgarSystematicObservationsSlip2017} analysed slip pulse behaviour and found a correlation between rise time and moment magnitude, making magnitude assessment possible after $\sim 15$~s.
While this conclusion would contradict our results, the significance of the findings for events with $M_w > 7.5$ is unclear, given the low number of very large events and several intermediate events with high rise time.
On the other hand, $\sim 15$~s does not imply any further predictability than found in our study for events with $M_w \leq 7.5$, due to their comparatively short duration.
Furthermore, the study by \citet{melgarSystematicObservationsSlip2017} uses geodetic observations in contrast to the STFs and teleseismic waveforms used in our analysis.
While teleseismic P arrivals should allow for good rupture tracking, similar to geodetic recordings, specific patterns of slip pulses might not be identifiable.

\citet{colombelliEarlyRuptureSignals2020} found differences in the slope of early peak ground motion parameters at local distances between earthquakes of magnitude 4 to 9.
In our analysis of teleseismic waves, this effect could be hidden by the attenuation of high frequency waveforms.
Therefore, our results do neither confirm nor contradict \citet{colombelliEarlyRuptureSignals2020}.
Similarly, while our study practically rules out predictability given STFs and teleseismic waveforms, it still leaves the option of approaches, where the tell-tale signals might only be observable in local waveforms, or require geodetic observations.

\section{Conclusion}

We conclude that there are no signs of early rupture predictability in either STFs or broadband teleseismic P waveforms.
Instead, our analysis indicates that the total moment of an event based on such data can only be estimated after the peak moment release.
However, even then it is not possible to anticipate future asperities.

While our analysis finds no early predictability, it highlights the feasibility of real time rupture tracking, at least using STFs and teleseismic waveforms.
Transferring the methods to regional waveforms might therefore still significantly benefit early warning.
Care has to be taken to counteract potentially biased estimations for the largest events, for which undersampling is hard to avoid \citep{munchmeyerTransformerEarthquakeAlerting2020}.

\section*{Data availability}
Moment rate datasets were obtained from the US Geological Survey (\url{https://earthquake.usgs.gov/data/finitefault/}), Linling Ye (slip models are supplement to \citet{yeRuptureCharacteristicsMajor2016} and linked in the acknowledgements therein), and the SCARDEC project (\url{http://scardec.projects.sismo.ipgp.fr/}).
We downloaded manual phase picks from the ISC \citep{iscbulletin} and USGS \citep{usgs2017comcat}.
Seismic waveforms were downloaded from the IRIS and GEOFON data centers.
We use waveforms from the GE \citep{networkGE}, G \citep{networkG}, GT \citep{networkGT}, IC \citep{networkIC}, II \citep{networkII}, and IU \citep{networkIU} seismic networks.

\section*{Acknowledgements}

Jannes Münchmeyer acknowledges the support of the Helmholtz Einstein International Berlin Research School in Data Science (HEIBRiDS). We thank Martin Vallée for the insightful discussion regarding the apparent early predictability in the SCARDEC dataset.

\bibliographystyle{gji}
\bibliography{bibliography,TeleMag}  %

\clearpage
\appendix

\clearpage

\renewcommand\thefigure{S\arabic{figure}}    
\setcounter{figure}{0}    

\renewcommand\thetable{S\arabic{table}}    
\setcounter{table}{0}  

\section{Apparent early predictability in SCARDEC}
\label{sec:artifacts}

To identify the source of the different behavior of the model between the STF datasets, we analyze the cumulative and current moment release at fixed early times and after a fixed moment release (Figure \ref{fig:artifacts}).
For SCARDEC, cumulative and current moment release at early times differs between different magnitude bins, with higher magnitude events already exhibiting higher moment release.
The same, even though with higher overlap between bins, is true for the current moment release at the time when magnitude 6 is reached.
No difference between bins are visible at the times when magnitudes 6.5 and 7 are reached.
These observations match the predictive results, where predicted magnitudes differed for early times and the time of reaching magnitude 6, but not for higher base magnitudes.

Differently from SCARDEC, for the other two STF datasets no systematic difference in any of the observables between the magnitude bins is visible.
This matches the predictive results, where predictions did not show systematic differences between buckets.
Given these observations and the processing of SCARDEC, in particular the point source approximation, we attribute the difference in the early observables to a processing artifact rather than interpreting them to be physically based.
In particular, the difference can be explained with uncertainties in the onset times for the SCARDEC STFs.
SCARDEC onset times are defined by the first time the STF exceeds a few percent of the peak moment rate.
This is necessary, as for an event with peak moment rate $>10^{20}$~Nm/s it will be impossible to identify the first exceedance of a low threshold such as $10^{17}$~Nm/s due to model approximations, in particular the point source approximation.
On the other hand, for an event with peak moment rate $\sim 10^{18}$~Nm/s, this first exceedance is easy to determine.
This introduces a systematic bias in the first seconds \cite{valleeNewDatabaseSource2016}.
This bias has also been analyzed quantitatively in prior publications \cite{meierApparentEarthquakeRupture2020}.

As a further validation, we trained a model on the USGS dataset, which is the larger of the two finite fault solution based datasets.
The results confirm, that no signs of rupture determinism are visible (Figure \ref{fig:results_stf_usgs}).
On both, the USGS data it was trained on and on the other two data sets, no systematic difference in the predictions between different magnitude buckets is observable until at least half of the time has passed or half of the moment has been released.
Note that, due to the different marginal distribution of magnitudes in the USGS dataset compared to SCARDEC, early estimates are considerably higher than for the SCARDEC model.
In addition, the smallest SCARDEC events are systematically overestimated.
This behaviour is expected, as neural networks are usually unable to extrapolate.

\section{STF model and training}
\label{sec:stf_training}

In this section, we provide details on the training of the STF model.
We train the model using a continuous ranked probability score (CRPS).
The CRPS is defined as
\begin{align}
 CRPS(F, x) = - \int_{- \infty}^{\infty} (F(y) - \mathbbm{1}_{\{y \geq x\}})^2 dy
 \label{eq:crps_definition}
\end{align}
with $F$ the cumulative distribution function of the predicted probability, $x$ the observed value and $\mathbbm{1}$ the indicator function, being 1 for $y \geq x$ (in our case) and 0 otherwise.
The CRPS measures the distance in probability mass between true and predicted cumulative distribution functions \citep{matheson1976scoring}, and not only takes into account the prediction at the observed value as the more common log-likelihood.
This is particularly useful for gradient-based optimisation in face of the highly skewed GR prior distribution.
The CRPS of a Gaussian mixture has a closed form representation and is differentiable with respect to the mixture parameters, making it amendable to gradient based optimisation (Text S\ref{sec:crps}).

As model, we use a multi-layer perceptron with five hidden layers with 200 neurons each and ReLU activation.
As input we use five observables derived from the source time function at time $t$: (1) cumulative moment $M_t$; (2) current moment rate $\dot{M}_t$; (3) average moment rate $\frac{1}{t} M_t$; (4) peak moment rate $\max_{\tau \leq t} \dot{M}_\tau$; (5) current moment acceleration $\ddot{M}_t$.
For improved learning behaviour, we log-transform features (1) to (4) and multiplied them by 0.1.
As feature (5) can take negative values as well, we transformed the feature with the function $f(x) = 0.01 \sign(x) \max(0, \log(|x| / (10^{15}\,\text{Nm}/\text{s}^2)))$, i.e., we apply a signed and scaled log-transform.
To mitigate slight differences in onset times we rebase the STF times such that the last sample with a moment rate below $10^{15}$~Nm is a $t=0$.

The network outputs mixture weights $\alpha_i$, mean values $\mu_i$ and standard deviations $\sigma_i$.
The probability density function (PDF) of the mixture is $f(x) = \sum_i \alpha_i \sigma_i^{-1} \varphi(\frac{x - \mu_i}{\sigma_i})$, where $\varphi$ denotes the PDF of a standard normal random variable.
For the mixture weights we use softmax activation, for the mean values no activation function and for the standard deviation softplus activation.
As we observed a mode collapse of the Gaussian mixture, i.e., all mixture components except one or two having mixture weights very close to zero, we introduced a Dirichlet prior on the mixture weights \cite{ormoneitImprovedGaussianMixture1995}, forcing mixture weights away from zero.
This regulariser takes the form $- \gamma \sum_i \log \alpha_i$.
For positive $\gamma$ this enforces that no mixture weight is close to zero.
We use $\gamma = 10^{-4}$.

We train the model using ten-fold cross validation with random splits.
In each split we use eight folds for training, one fold for validation and the last one as test set.
For each split we train five models and average the predictions in probability space.
We use the Adam optimiser with learning rate $10^{-4}$ and a batch size of 128.
We reduce the learning rate by a factor of 0.3 after 15 epochs without a reduction in validation loss.
We train for 100 epochs and use from each ensemble member the model with lowest validation loss for evaluation.

\citet{munchmeyerEarthquakeMagnitudeLocation2021} showed that neural network models suffer from data sparsity for large events, causing systematic underestimation of magnitudes.
As a simple mitigation, they suggested upsampling these events in training, i.e., artificially increasing their occurrence.
We follow this approach by upsampling events above magnitude 6 with the factor $\rho(M) = \lambda ^ {M - 6}$, where we use $\lambda = 2$.
As we analyse probabilistic predictions, we need to take the introduced skew on the distribution into account.
For this, we analyse Bayes' rule $\mathbb{P}(M|O_t) \sim \mathbb{P}(O_t|M) \mathbb{P}(M)$.
The upsampling replaces $\mathbb{P}(M)$ by $\tilde{\mathbb{P}}(M) = c_1 \rho(M) \mathbb{P}(M)$.
The model therefore estimates $\tilde{\mathbb{P}}(M|O_t) = c_2 \rho(M) \mathbb{P}(M|O_t)$, where $c_1$ and $c_2$ are simply normalisation constants.
Note that usually $c_1 \neq c_2$, as the normalisation constant $\mathbb{P}(O_t)$ will change with the upsampling as well.
To get true estimates of $\mathbb{P}(M|O_t)$, one would need to rescale the predictions with $1 / \rho(M)$.
Notably, this scale factor is independent of $O_t$.

For the results presented in this thesis we refrained from rescaling the predictions for several reasons.
First, our upsampling rate of 2 per magnitude step is considerably weaker than the Gutenberg-Richter law with a tenfold decrease in event occurrence with each magnitude step. 
Therefore upsampling will not obscure Gutenberg-Richter tails.
In fact, the lower decay rate with magnitude obtained by upsampling allows for better visual representation.
Second, our key results compare predictions in different buckets.
As each bucket is equivalently affected by the upsampling, their relative behaviour stays unchanged.
However, we note that any quantitative analysis should take the effect of upsampling into account.

\section{Teleseismic arrival dataset and model}
\label{sec:predictability:teleseismic_data}

We downloaded all available manual phase picks for events with magnitudes above 5 from the ISC \citep{iscbulletin} and USGS \citep{usgs2017comcat}.
We matched the event references to the Global CMT catalog \citep{ekstromGlobalCMTProject2012} and discarded all events that could not be matched.
The station and event distribution is visualised in Figure \ref{fig:telemag_map}, alongside the magnitude and epicentral distance distributions.
For all analysis, we used the moment magnitude from Global CMT as target value.
We only use picks with phase label P and discarded all but the first pick for each station and event.
We only use picks within an epicentral distance below $97^\circ$ to avoid core phases.
For consistency, we calculated expected first P arrival times using the GCMT event onset times and the ak135 velocity model.
If a pick was not within 4~s of the first predicted arrival, we discarded the pick.
We use broadband waveforms from the \citep[GE,][]{networkGE}, \citep[G,][]{networkG}, \citep[GT,][]{networkGT}, \citep[IC,][]{networkIC}, \citep[II,][]{networkII} and \citep[IU,][]{networkIU} seismic networks that we downloaded from the GEOFON and IRIS FDSN webservices.
We excluded all stations within 10~km of the coastline as they showed high levels of short-period noise.
We do not enforce any further constraint on the signal to noise ratio, but note that the usage of manual picks provides an implicit constraint.
All waveforms are resampled to 20~Hz sampling rate, filtered between 0.025~Hz and 8~Hz and cut from 35~s before the phase pick to 90~s after the phase pick.
We removed the instrument sensitivity, but did not restitute the instrument response as we observed acausal artefacts from restitution.
We manually inspected the resulting data set and removed stations with timing errors.
As sanity check, we applied our model to $t=-0.5$~s, i.e., 0.5~s before the annotated P arrivals.
The results showed no significant difference from the marginal distribution of magnitudes, indicating no or at least very few cases with severe timing errors or other knowledge leaks.
The resulting catalog consists of 37,646 events with 747,824 manually labelled P arrivals from 307 unique seismic stations.

As neural network we adapted TEAM-LM \cite{munchmeyerEarthquakeMagnitudeLocation2021}.
Compared to the original publication, we introduced several modifications to TEAM-LM to fit our application.
First, as the traces are teleseismic, it was not possible to align the traces between stations by wall time.
Instead, we align the traces by their P picks, such that the P pick is at the same sample for each station.
Second, we now model real-time application through a sliding window instead of zero padding.
To model the data available at time $t$, where $t$ is relative to the P pick, we provide the model with the waveforms from $t-30$~s to $t$.
This allows to (i) apply the model to times more than $30$~s after the P arrival;
(ii) give the model more information of the noise at early times; (iii) make the model less sensitive to pick inaccuracies.
Third, we do not encode station positions.
We experimented with encoding the positions relative to the event, but it became apparent that the station distribution in our dataset in many cases is indicative of the magnitude.
However, at teleseismic distances, the locations generally tend to have lower impact on the waveform than at regional distances, which is also visible for our results.
In addition, we modified the mixture density output to be consistent with the one for the STF model, i.e., we added the Dirichlet regularization and switched to softplus for the sigma values.

We train the model using 10 fold cross validation with random splits.
In each split we use 8 folds for training, 1 fold for validation and the last one as test set.
Due to the massively higher computing requirements for the TEAM-LM model compared to the STF model, we did not train an ensemble but only a single model for each split.
We use the Adam optimiser with learning rate $10^{-4}$ and a batch size of 1024.
We reduce the learning rate by a factor of 0.3 after 5 epochs without a reduction in validation loss.
We train for 100 epochs and use the model with lowest validation loss for evaluation.
We clip gradients to a maximum norm of 1.
We use at most 50 input stations.
As for the STF model we use upsampling of large magnitude events and did not rescale the outputs.
As \citet{munchmeyerEarthquakeMagnitudeLocation2021}, we pretrain the feature extraction and the mixture density layers on single station magnitude estimation.
As the extensive data augmentation incorporates stochasticity in the validation score, the validation set is evaluated five times with different augmentations after each epoch.

\section{CRPS of Gaussian mixture}
\label{sec:crps}

Here we derive the closed form solution of the CRPS for a Gaussian mixture.
The probability density function (PDF) of a Gaussian mixture is defined as $f(x) = \sum_i \alpha_i \sigma_i^{-1} \varphi(\frac{x - \mu_i}{\sigma_i})$, where $\varphi$ denotes the PDF of a standard normal random variable.
Similarly we use $\varPhi$ for the cumulative distribution function (CDF) of a standard normal random variable.
For deriving the closed form solution, we use three identities.
First, Gneiting and Raftery \cite{gneitingStrictlyProperScoring2007} note that the CPRS (eq.~1) can be written as
\begin{align}
 CRPS(F, x) = \frac{1}{2} \mathbb{E}_F|X - X'| - \mathbb{E}_F|X - x|  \label{eq:crps}
\end{align}
with $X$ and $X'$ independent copies of random variables with CDF $F$, and $\mathbb{E}_F|\cdotp|$ being the expectation of the absolute value.
Note that this identity requires a finite first moment of $X$, which for a finite Gaussian mixture is always true.
Second, for two independent Gaussian random variables $Z \sim \mathscr{N}(\mu, \sigma^2)$ and $Z' \sim \mathscr{N}(\mu', \sigma'^2)$, the sum $Z + Z'$ is a Gaussian random variable with $Z + Z' \sim \mathscr{N}(\mu + \mu', \sigma^2 + \sigma'^2)$.
Third, for a Gaussian random variable $Z \sim \mathscr{N}(\mu, \sigma^2)$ the expected absolute value has the following closed form solution:
\begin{align}
 \mathbb{E}|Z| = 2 \sigma ^ 2 \varphi \left( \frac{\mu}{\sigma} \right) + \mu \left( 1 - 2 \varPhi \left( - \frac{\mu}{\sigma}\right) \right) \label{eq:exp_fold_normal}
\end{align}

Let $X$ and $X'$ be the Gaussian mixtures and $Z_i, Z'_i \sim \mathscr{N}(\mu_i, \sigma_i^2)$ be the mixture components.
We can now calculate the terms of (\ref{eq:crps}).
For the first term we get:
\begin{align}
 \mathbb{E}|X - X'| &= \sum_{i=1}^n \sum_{j=1}^n \alpha_i \alpha_j \mathbb{E}|Z_i - Z'_j| \\
                    &= \sum_{i=1}^n \sum_{j=1}^n \alpha_i \alpha_j \mathbb{E}|N_{ij}|
\end{align}
Here, $N_{ij} \sim \mathscr{N}(\mu_i - \mu_j, \sigma_i^2 + \sigma_j^2)$, using the summation of independent Gaussian random variables.
The expected value can be computed using (\ref{eq:exp_fold_normal}).

Similarly, for the second term of (\ref{eq:crps}) we get
\begin{align}
 \mathbb{E}|X - x| &= \sum_{i=1}^n \alpha_i \mathbb{E} |Z_i - x| \\
                   &= \sum_{i=1}^n \alpha_i \mathbb{E} |Y_i|
\end{align}
with $Y_i \sim \mathscr{N}(\mu_i - x, \sigma_i)$.
This again allows to calculate the term using (\ref{eq:exp_fold_normal}).
Therefore, the CRPS of the Gaussian mixture can be computed in closed form.
Furthermore, the solution is differentiable in $\alpha_i$, $\mu_i$ and $\sigma_i$, which is required for neural network training.
For $\alpha_i$ differentiability is clear, as the CRPS only depends linearly on the mixture weights.
For $\mu_i$ and $\sigma_i$, differentiability results from the differentiability of $\varphi$ and $\varPhi$.
While this does not hold true for $\sigma_i = 0$, our network architectures ensure $\sigma_i > 0$.
Calculating the closed form has compute complexity in $\mathcal{O}(n^2)$.
As the number of mixture components is low ($n < 25$) and the calculation can trivially be vectorized, this does not pose a computational issue and computation times are negligible compared to the neural network computations.

\clearpage

\begin{figure}
 \includegraphics[width=\textwidth]{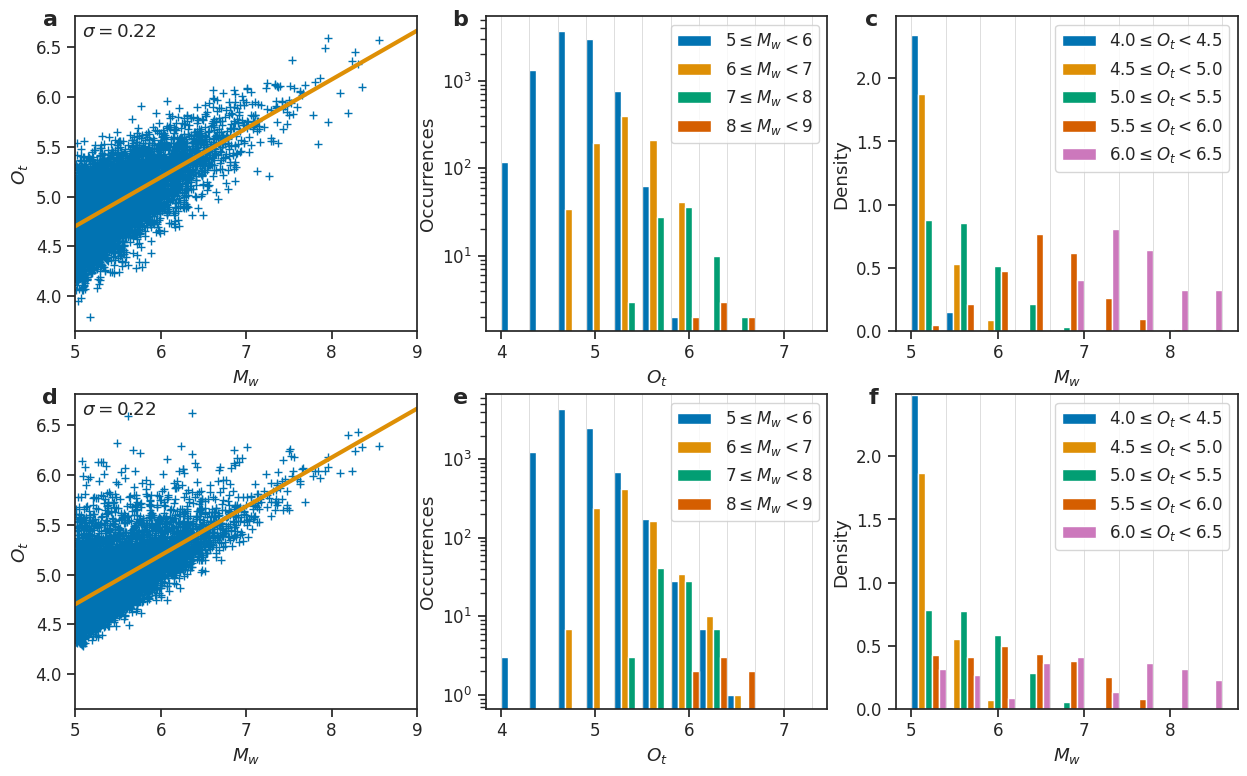}
 \caption{\footnotesize{Synthetic samples of an arbitrary scalar observable $O_t$ and magnitude $M_w$ assuming a linear connection with Gaussian error (\textbf{a-c}) or with exponentially modified Gaussian error (\textbf{d-f}), i.e., the sum of a Gaussian and an exponential random variable. The parameters of the modified Gaussian distribution were chosen such that its standard deviation is the same as in the first case. For both cases large magnitude events cause large observables, but the contrary is only true for the Gaussian case, i.e., in the second case small events can cause large observables as well. \textbf{a} and \textbf{d} show scatter plots of $O_t$ and $M_w$. \textbf{b} and \textbf{e} show histograms of $O_t$ for $M_w$ bins with log-scaled y axis. The observable distributions for different magnitudes are mostly distinct in the first case, but overlap strongly for the second case. \textbf{c} and \textbf{f} show histograms of $M_w$ for $O_t$ bins which are normed to represent densities. The magnitude distributions for different observables are mostly distinct for the first case, while for the second case, observables only give an upper bound on the magnitude. $M_w$ samples were generated according to a Gutenberg-Richter distribution with $b=1$. $O_t$ samples were generated using the linear connection and random samples from the error distribution. Both models assume the same linear connection.}}
 \label{fig:probabilistic_errors}
\end{figure}

\begin{figure}
 \includegraphics[width=\textwidth]{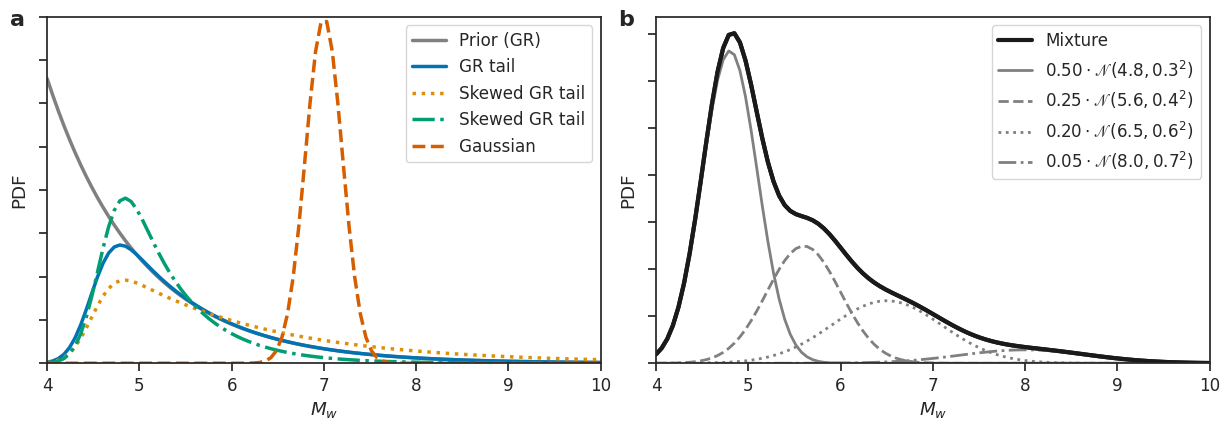}
 \caption{\textbf{a} Possible shapes of $\mathbb{P}(M|O_t)$ for an ongoing event. The Gutenberg-Richter prior is rescaled to fit the tail behavior of the other distributions.
 \textbf{b} Exemplary Gaussian mixture with mixture size 4, showing both the individual components and the resulting mixture PDF.}
 \label{fig:pmo_shapes_mixture}
\end{figure}

\begin{figure}
 \includegraphics[width=\textwidth]{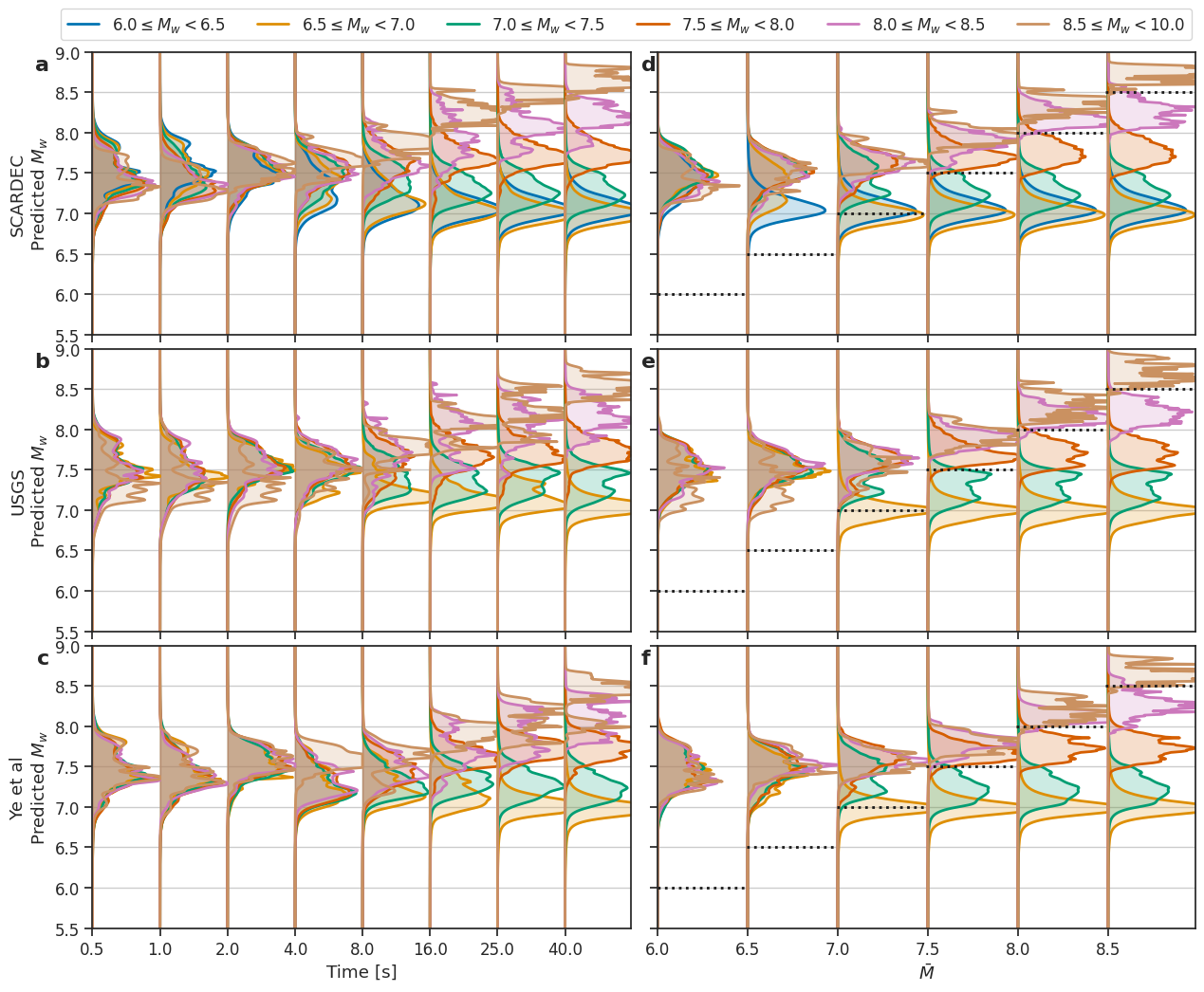}
 \caption{Results similar to Figure 4, but using the USGS STFs instead of the SCARDEC ones for model training. For details see the description of Figure 4. Note that the marginal distribution of magnitudes in the USGS dataset is considerably different from the SCARDEC dataset, i.e., it is missing smaller events. This is clearly reflected in the results, in particular in the overestimation of small SCARDEC events.}
 \label{fig:results_stf_usgs}
\end{figure}

\begin{figure}
 \includegraphics[width=\textwidth]{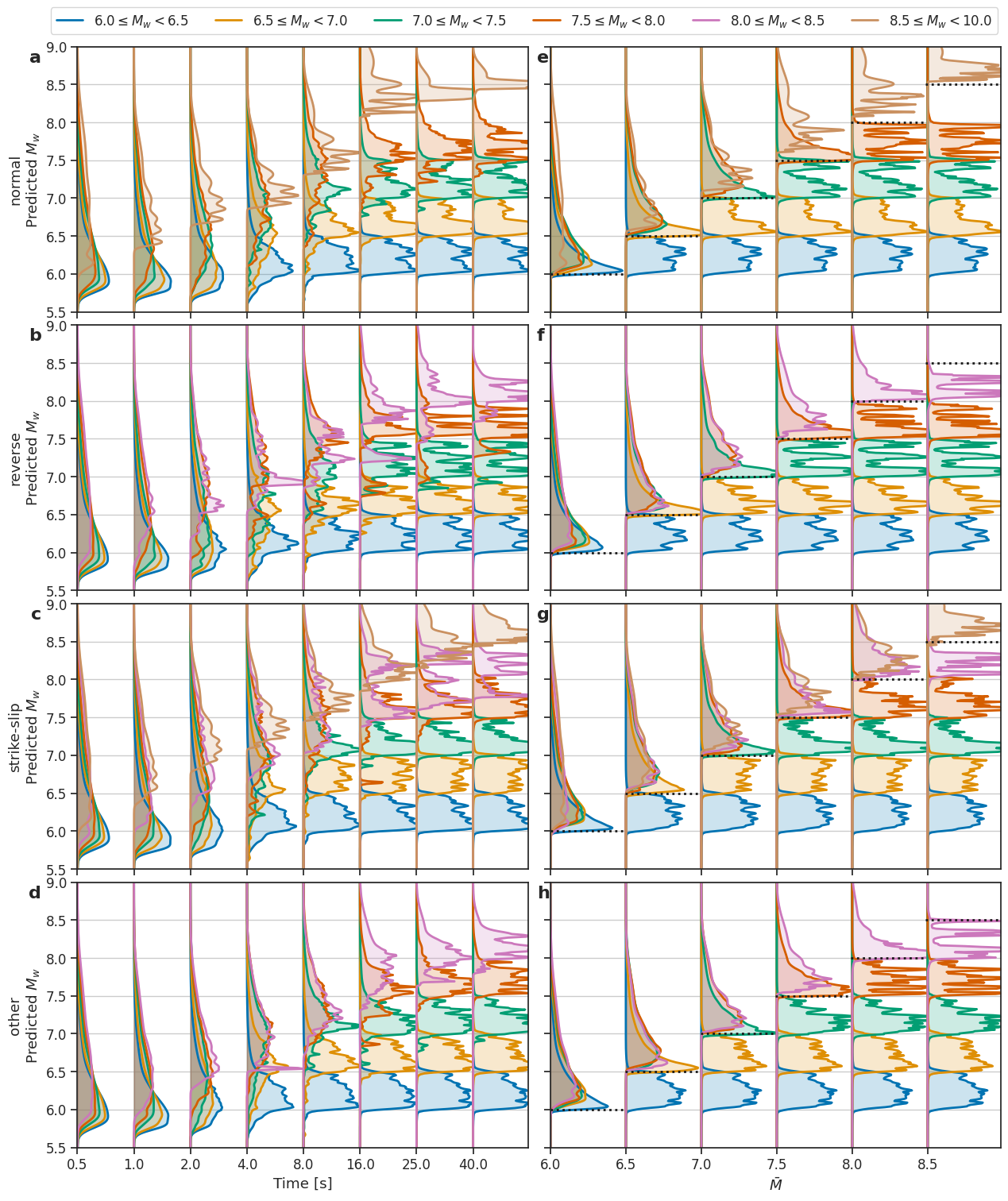}
 \caption{\footnotesize{Average $\mathbb{P}(M|O_t)$ (\textbf{a-d}) and $\mathbb{P}(M|O_{\bar{M}})$ (\textbf{e-h}) by magnitude bin for the SCARDEC dataset. This figure displays the same results as shown in Figure 4a, b but with the analysis split by focal mechanism type. 
 Focal mechanism types were derived from the Global CMT solution using the principal axis. If the $n$ axis was within 30$^\circ$ of the horizontal, the event was classified as ``normal'' ($t$ axis more vertical than $p$ axis)  or ``reverse'' ($p$ axis more vertical than $t$ axis). If the $n$ axis was within 30$^\circ$ or the vertical axis, the event was classified as ``strike-slip''. All remaining events were classified as ``other''.
 PDFs were truncated to avoid overlap between different times/base magnitudes. Black dotted lines in \textbf{e-h} indicate the current base magnitude. For events shorter than the given time (\textbf{a-d}) or with final magnitudes below the base magnitude (\textbf{e-h}), the estimation from the final sample of the STF was used.}}
 \label{fig:results_stf_eqtypes_scardec}
\end{figure}

\begin{figure}
 \includegraphics[width=\textwidth]{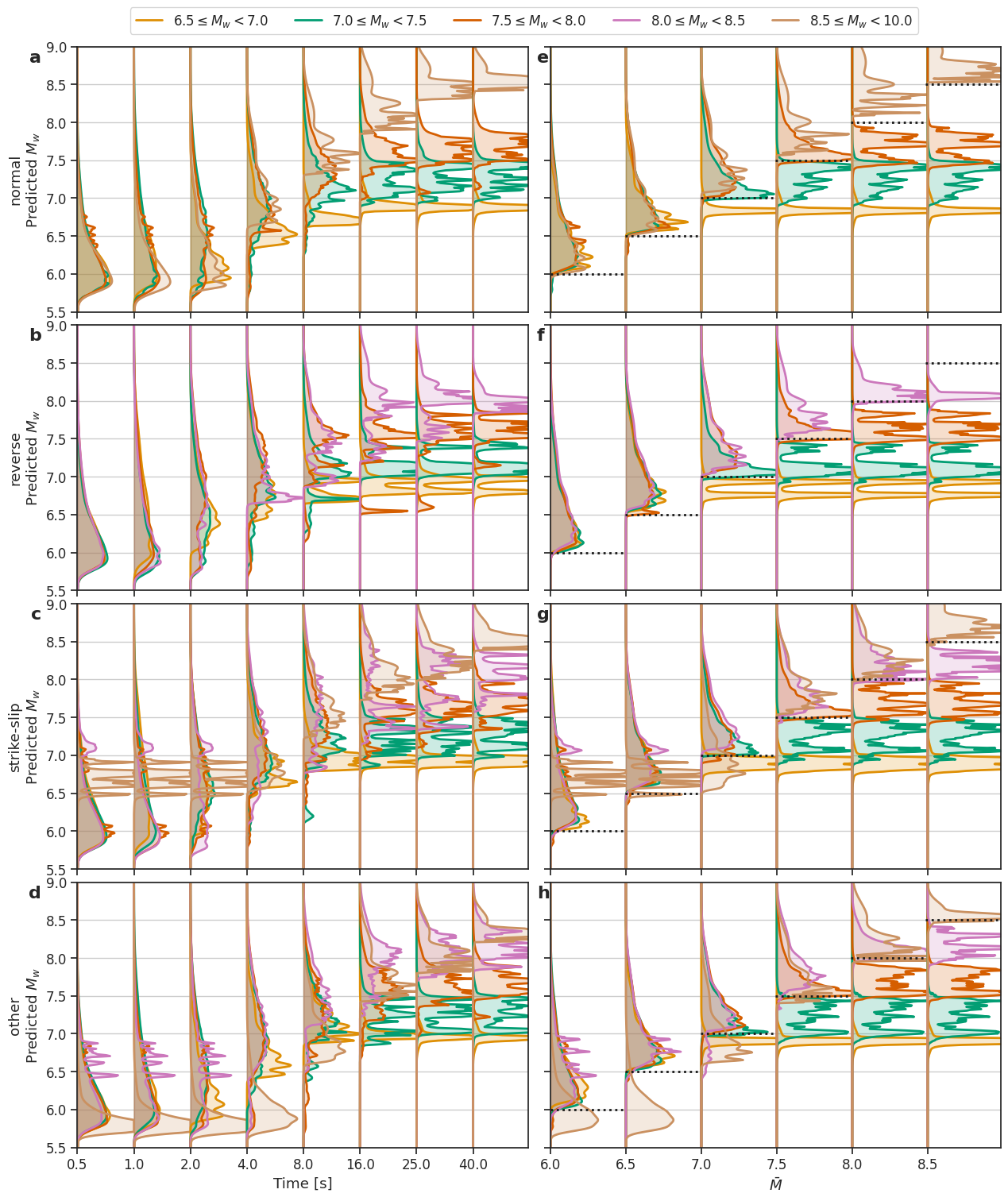}
 \caption{Average $\mathbb{P}(M|O_t)$ (\textbf{a-d}) and $\mathbb{P}(M|O_{\bar{M}})$ (\textbf{e-h}) by magnitude bin for the USGS dataset. 
 This figure displays the same results as shown in Figure 4c, d but with the analysis split by focal mechanism type.
 Otherwise, see caption of Figure~\ref{fig:results_stf_eqtypes_scardec} for further explanations. }
 \label{fig:results_stf_eqtypes_usgs}
\end{figure}

\begin{figure}
 \includegraphics[width=\textwidth]{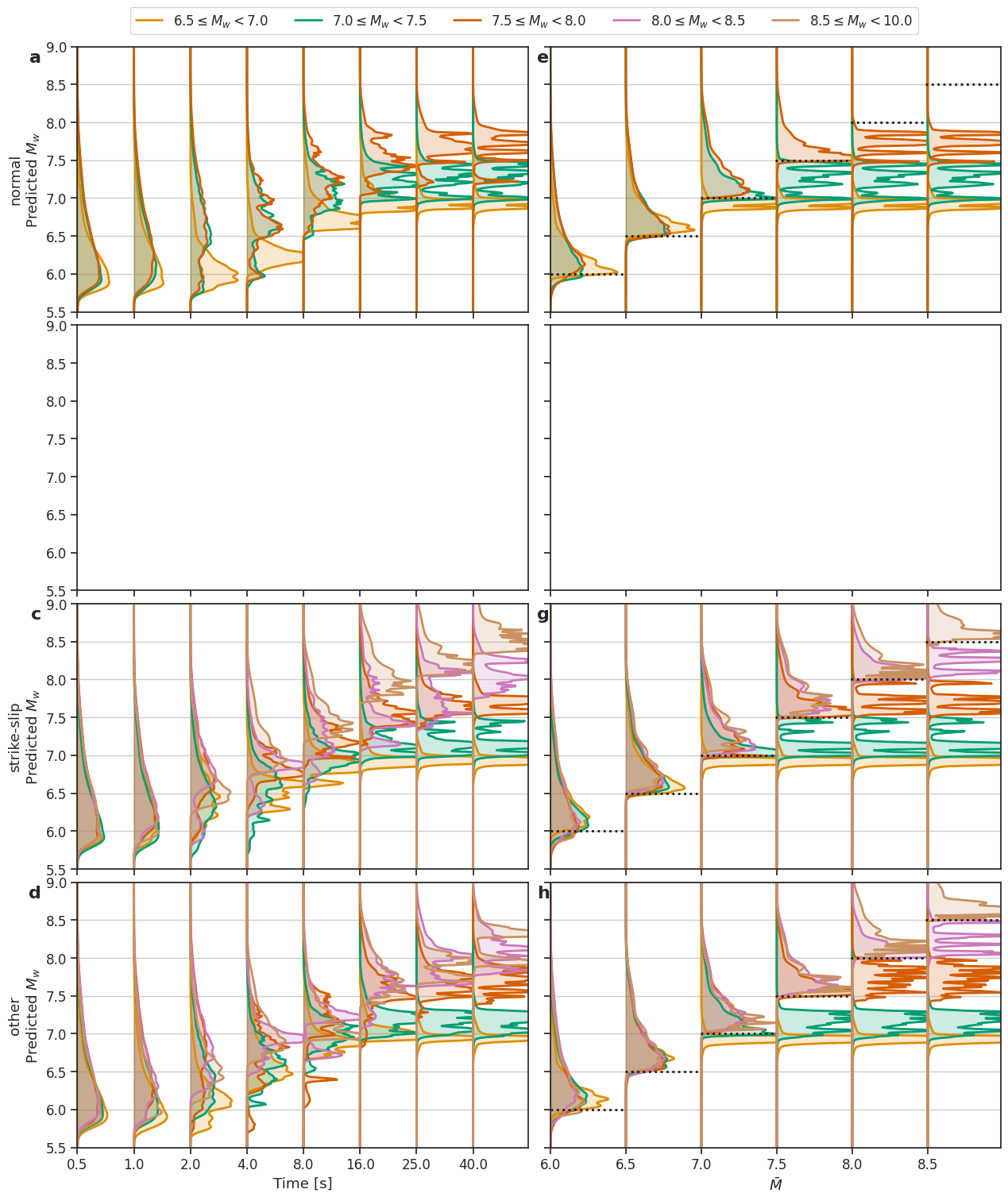}
 \caption{Average $\mathbb{P}(M|O_t)$ (\textbf{a-d}) and $\mathbb{P}(M|O_{\bar{M}})$ (\textbf{e-h}) by magnitude bin for the Ye et al dataset. 
 This figure displays the same results as shown in Figure 4e, f but with the analysis split by focal mechanism type.
 The dataset contains no examples of reverse faulting, therefore the corresponding panels are left empty.
 Otherwise, see caption of Figure~\ref{fig:results_stf_eqtypes_scardec} for further explanations.}
 \label{fig:results_stf_eqtypes_yeetal}
\end{figure}

\begin{figure}
 \includegraphics[width=0.5\textwidth]{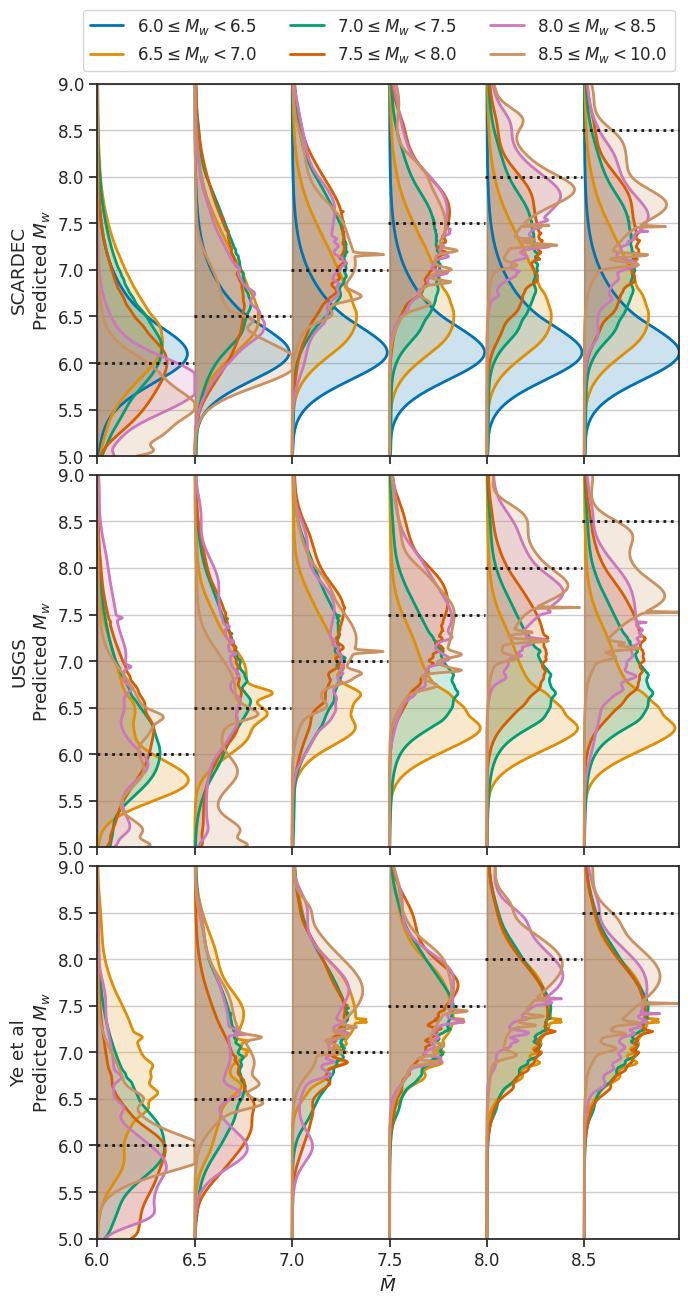}
 \caption{$\mathbb{P}(M|O_{\bar{M}})$ binned by magnitude using the three STF datasets for determining $t_{\bar{M}}$ in \textbf{b}. The figure is otherwise equivalent to Figure 4h; for more details of the figures format also see the caption of Figure 4. Note that the events shown differ between the panels, as only those events included in the respective STF datasets can be shown.}
 \label{fig:results_team_datasets}
\end{figure}

\begin{figure}
 \includegraphics[width=\textwidth]{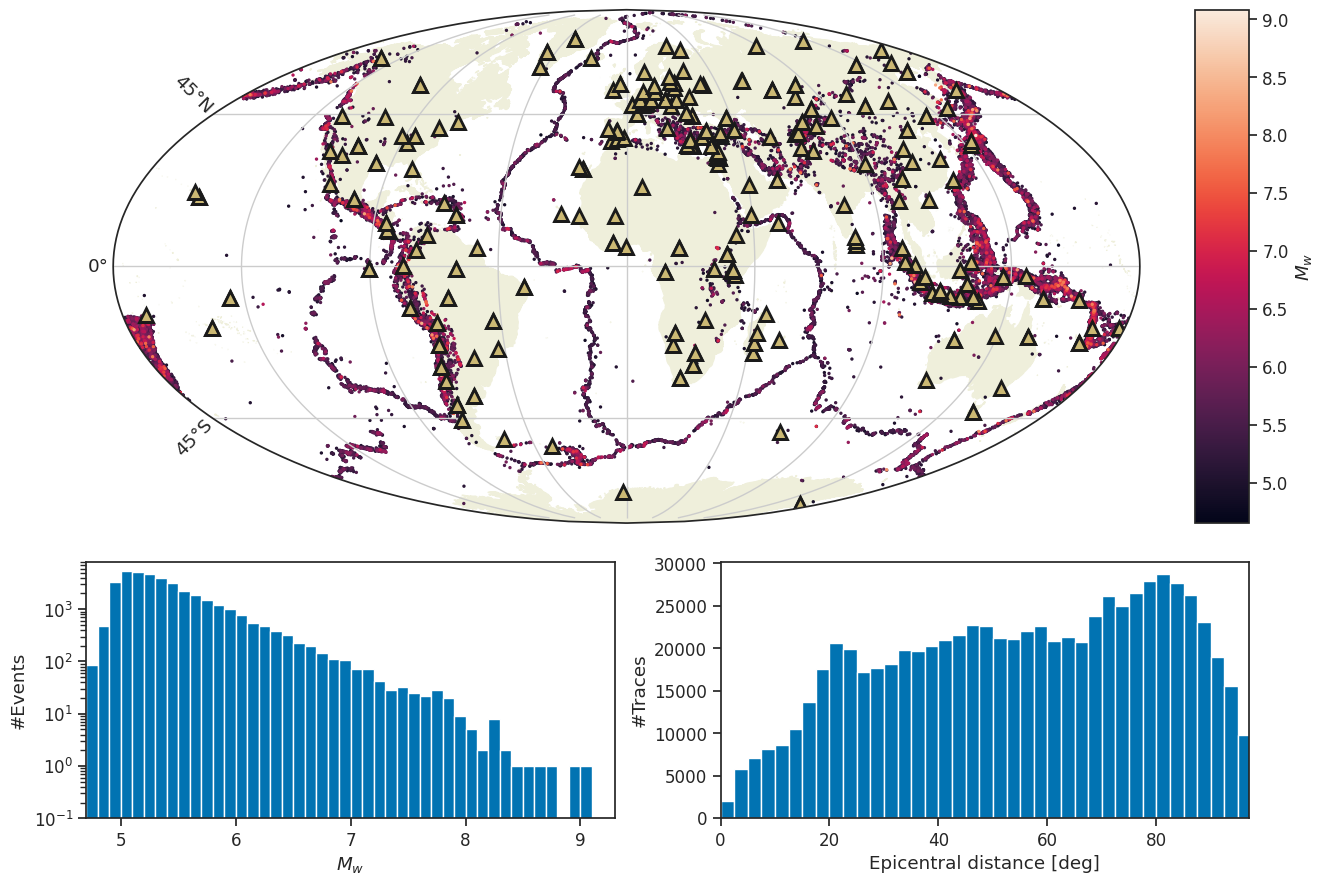}
 \caption{Distribution of stations and events and histograms for magnitude and epicentral distance distributions for teleseismic P arrival dataset. In the map, triangles denote stations and dots denote events. Events are color-coded by magnitude.}
 \label{fig:telemag_map}
\end{figure}

\begin{figure}
 \includegraphics[width=\textwidth]{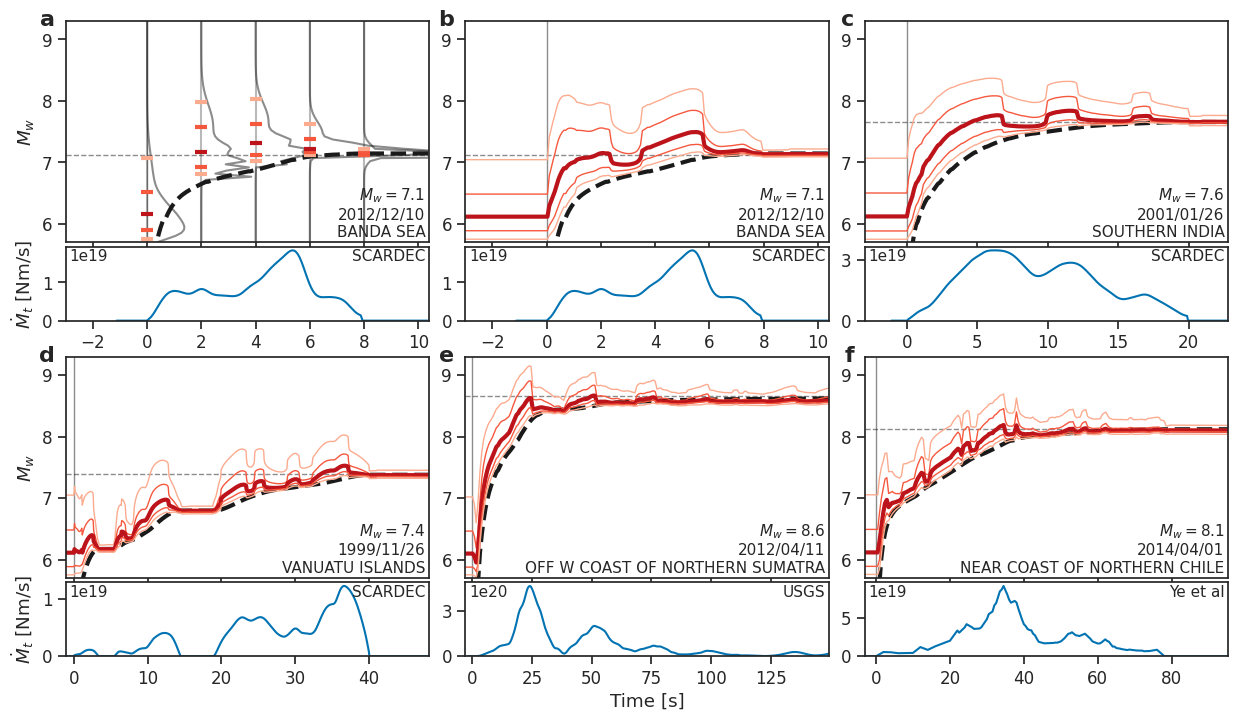}
 \caption{Results similar to Figure 3, but using the optimal SCARDEC STFs instead of the average ones for model training and evaluation on SCARDEC. For details see the description of Figure 3.}
 \label{fig:results_stf_anecdotal_optimal}
\end{figure}

\begin{figure}
 \includegraphics[width=\textwidth]{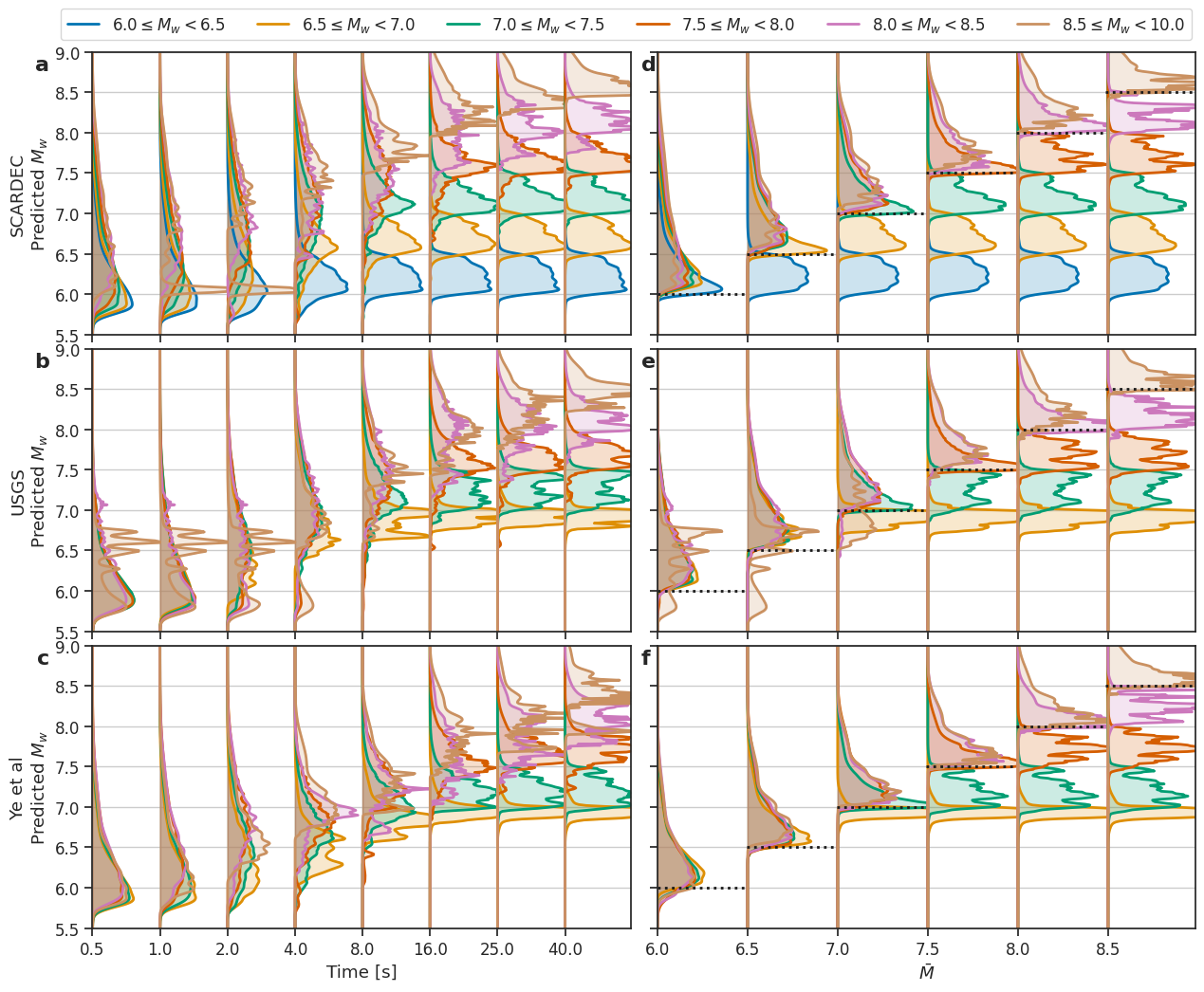}
 \caption{Results similar to Figure 4, but using the optimal SCARDEC STFs instead of the average ones for model training and evaluation on SCARDEC. For details see the description of Figure 4.}
 \label{fig:results_stf_optimal}
\end{figure}

\begin{figure}
 \includegraphics[width=\textwidth]{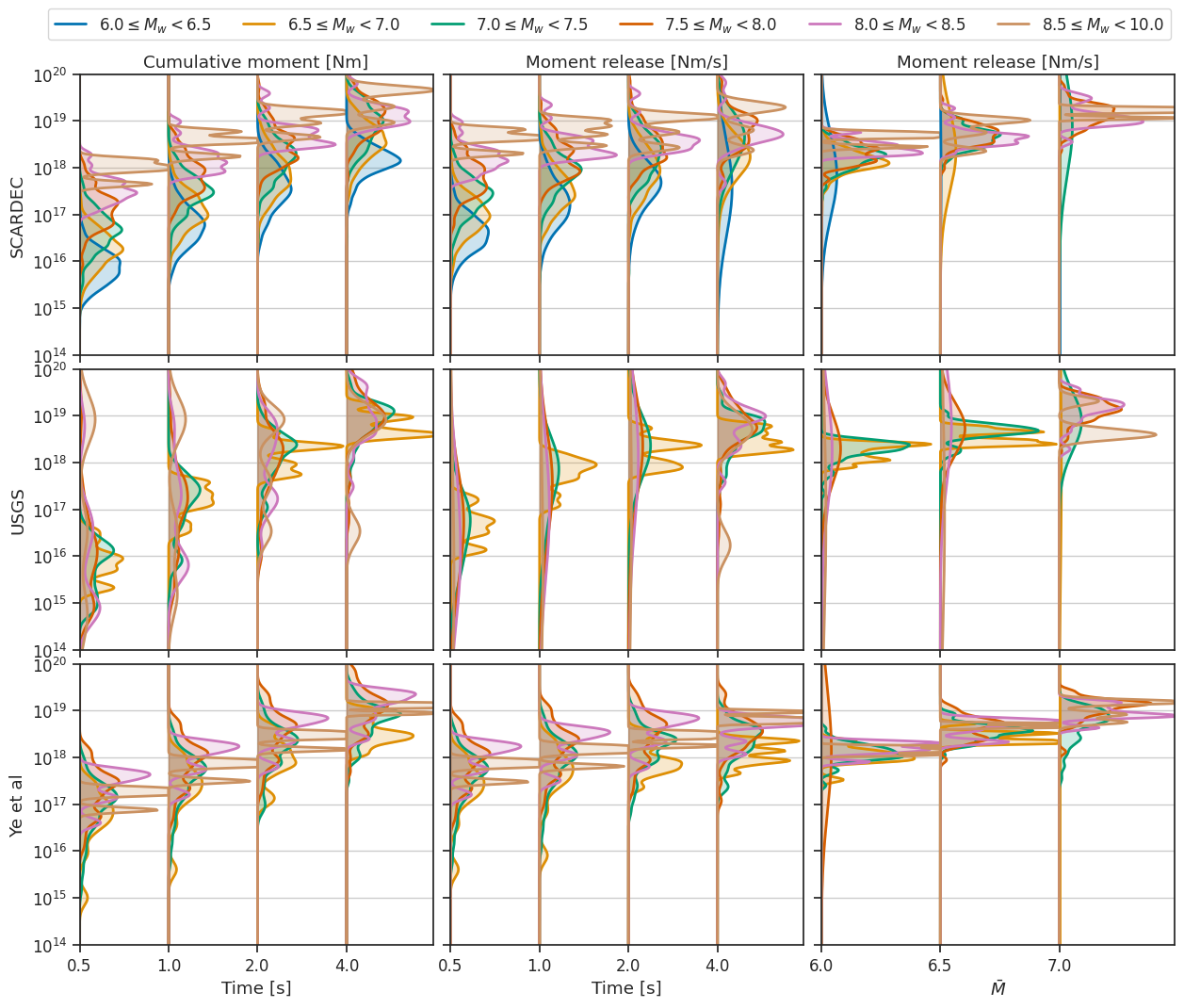}
 \caption{Comparative analysis of the early moment release for the three STF datasets binned by magnitude. Each row represents one STF dataset. The left column shows cumulative moment release at time $t$, the middle column current moment release at time $t$, the right column moment release at the time when a magnitude $\bar{M}$ is reached. Notably, while all three measures differ between the magnitude buckets for SCARDEC, no such behavior is visible for the USGS or Ye et al datasets. This points at a processing artifact in SCARDEC rather than a physical explanation.}
 \label{fig:artifacts}
\end{figure}

\end{document}